\documentclass{aa}
\usepackage{graphicx,float}
\begin{document}

\title{Characterising anomalous transport in accretion disks from X-ray observations}

\author{J.~Greenhough\inst{1}\and S.~C.~Chapman\inst{1}\and S.~Chaty\inst{2}\and R.~O.~Dendy\inst{3,1}\and G.~Rowlands\inst{1}}

\offprints{J.~Greenhough}

\institute{Department of Physics, University of Warwick, Coventry CV4 7AL, UK\\
\email{\{greenh, sandrac\}@astro.warwick.ac.uk}
\email{g.rowlands@warwick.ac.uk}
\and Department of Physics \& Astronomy, The Open University, Milton Keynes MK7 6AA, UK\\
\email{S.Chaty@open.ac.uk}
\and Euratom/UKAEA Fusion Association, Culham Science Centre, Abingdon, Oxfordshire, OX14 3DB, UK\\
\email{richard.dendy@ukaea.org.uk}}

\date{Received 2001 July 4/ Accepted }

\abstract{
Whilst direct observations of internal transport in accretion disks are not yet possible, measurement of the energy emitted from accreting astrophysical systems can provide useful information on the physical mechanisms at work. Here we examine the unbroken multi-year time variation of the total X-ray flux from three sources: \object{Cygnus X-1}, the microquasar \object{GRS1915+105}, and for comparison the nonaccreting \object{Crab} nebula. To complement previous analyses, we demonstrate that the application of advanced statistical methods to these observational time-series reveals important contrasts in the nature and scaling properties of the transport processes operating within these sources. We find the Crab signal resembles Gaussian noise; the Cygnus X-1 signal is a leptokurtic random walk whose self-similar properties persist on timescales up to three years; and the GRS1915+105 signal is similar to that from Cygnus X-1, but with self-similarity extending possibly to only a few days. This evidence of self-similarity provides a robust quantitative characterisation of anomalous transport occuring within the systems. 
\keywords{accretion disk -- methods: statistical -- X-rays: individual (Crab, Cygnus X-1, GRS1915+105)}
}

\titlerunning{Anomalous transport in accretion disks}
\maketitle

\section{Introduction}
Deeper understanding of the transport mechanisms that operate within accretion disks is important for a broad range of X-ray emitting astrophysical objects. In the absence of local measurements, the key questions are (1)~to quantify the ways in which the unseen transport processes are anomalous as distinct from diffusive; and (2)~to establish how this may be determined remotely from observations of global quantities such as the input and outflow of energy. This information can then be used to provide a constraint for turbulence/instability models of astrophysical accretion disks. 

\begin{figure}

	\resizebox{\hsize}{!}{\includegraphics{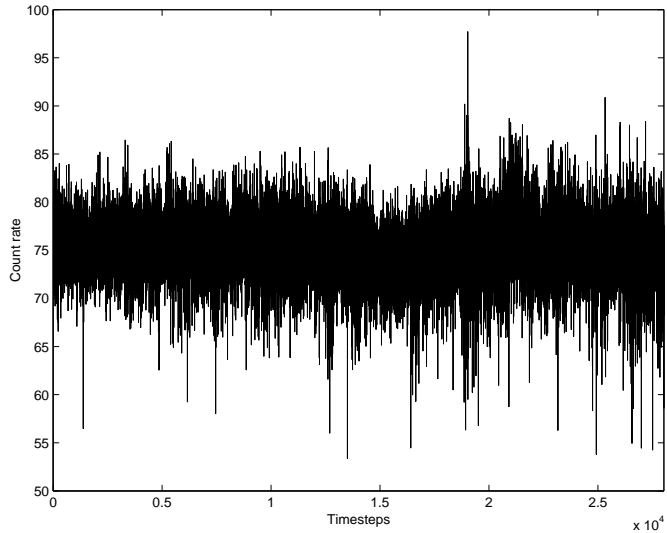}}
	\caption{X-ray time-series for the Crab, 1996 Feb. -- 2001 Mar; mean timestep 93 minutes.}\label{lccr}
\end{figure}

\begin{figure}

	\resizebox{\hsize}{!}{\includegraphics{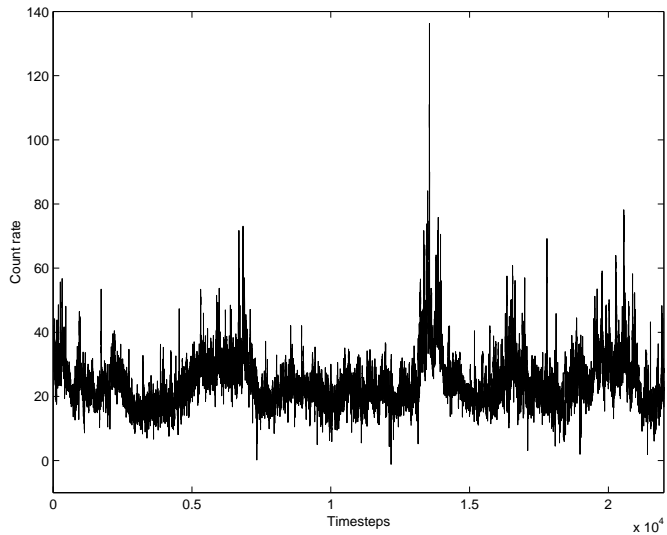}}
	\caption{X-ray time-series for Cygnus X-1, 1996 Sep. -- 1999 Dec; mean timestep 77 minutes.}\label{lccy}
\end{figure}

\begin{figure}

	\resizebox{\hsize}{!}{\includegraphics{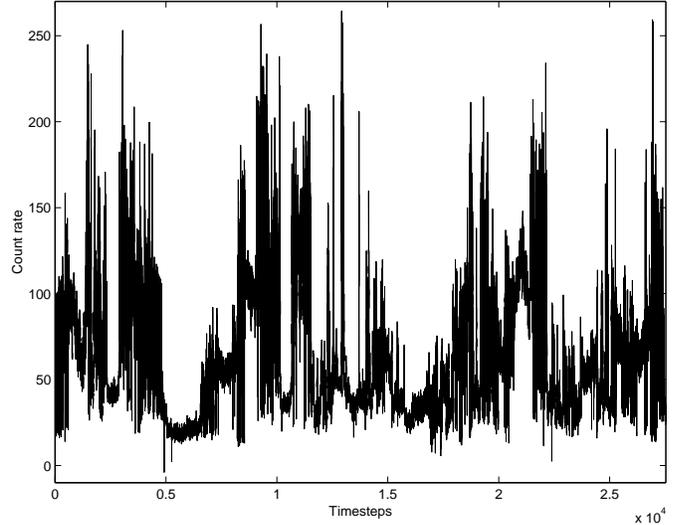}}
	\caption{X-ray time-series for GRS1915+105, 1996 Feb. -- 2001 Mar; mean timestep 96 minutes.}\label{lcgr}
\end{figure}

Here we analyse the total X-ray flux over several years from two accreting astrophysical objects -- Cygnus X-1 and GRS1915+105, the stellar-mass analogues of disk-jet active galaxies powered by a massive black hole -- and, for comparison, the nonaccreting Crab which is powered by a neutron star. The reasons for selecting these sources for statistical analysis are twofold. First, since 1996 February 20 they have been observed continually for several years by the All-Sky Monitor (ASM) on board the RXTE satellite (Swank et al. \cite{rxte}), providing large data sets of around thirty thousand points. This enables us to seek correlations over several orders of magnitude up to the longest accessible timescales. Second, the source luminosities are sufficiently high to neglect instrument thresholds, uncertainties, and other sources in the field of view. The raw data are held at the Goddard Space Flight Center (GSFC) and can be accessed via their website\footnote{http://heasarc.gsfc.nasa.gov/docs/xte/asm\_products.html}. Each point represents the total X-ray flux (measured by the number of counts during periods that last 90 seconds) in the range 1.3--12.2 $keV$, and the breakdown into three energy bands (1.3--3, 3--5 and 5--12.2 $keV$) is also available. We have analysed the different channels and found them to be similar to the total flux. Sampling intervals between the 90-second X-ray counting periods are distributed with means of 93 minutes for the Crab, 77 minutes for Cygnus X-1, and 96 minutes for GRS1915+105. To give an indication of the spread of sampling intervals, 90\% of the intervals are below 186 minutes for the Crab, 187 minutes for Cygnus X-1, and 194 minutes for GRS1915+105. The implications for our techniques are explained where appropriate below (and see Appendix). Calibration is undertaken by the ASM/RXTE team and the processed data are freely accessible on their website\footnote{http://xte.mit.edu/XTE/asmlc/ASM.html} (Bradt et al. \cite{asm}). The RXTE counts are not directly proportional to luminosity, so the exact relationship between these two quantities would have to be accounted for in any model. However, this will affect only the nature of the PDFs and not the temporal correlation in the data. The three raw X-ray time-series are plotted in Figs.~\ref{lccr}, \ref{lccy} and \ref{lcgr}.     

Previous studies have focused on spatial and spectral structure, quasi-periodic
oscillations, and modelling of temporal variability: see Weisskopf et al. (\cite{cr}) and references therein for the Crab; Maccarone et al. (\cite{cy}) and references therein for Cygnus X-1; and Belloni et al. (\cite{belloni}) and Rao et al. (\cite{rao}) for summaries of the spectral and temporal analyses of GRS1915+105. Dhawan et al. (\cite{dhawan}) have studied the jet of this source, and the importance of microquasars in general is discussed by Mirabel and Rodr\'{\i}guez (\cite{mirabel}). Nayakshin et al. (\cite{nay}) model the gross variability of GRS1915+105 on all but the shortest timescales, and recent observations by Chaty et al. (\cite{chaty}) have been used to search for interactions with the surrounding interstellar medium. 
As a complement to these techniques, we here examine three key statistical measures (described further in Sect.~2) for each time-series. These are the probability density function (PDF), the growth of range (extent of statistical self-similarity), and the differenced form, which we then compare with the well-known signatures of Gaussian noise and random walks. In particular, we test for signatures of long-range correlations that are the hallmarks of turbulent transport. Insofar as these techniques can be applied successfully to the observational signatures of anomalous transport in the present context, they may also be transferable to related questions in space and laboratory plasma physics.

There is already much evidence that accretion disks are locales for turbulent transport and instabilities. In order to explain typical accretion rates observed in a range of disk types, the standard disk model (Shakura \& Sunyaev \cite{shak}; Pringle \& Rees \cite{pringle}) uses turbulent viscosity to produce an appropriate outward transport of angular momentum. Following extensive numerical simulations (reviewed by, for example, Gammie \cite{gamm}), the source of this turbulence is now believed to be a magnetic shear instability (Balbus \& Hawley \cite{balb}). In addition, $1/f$ power spectra (inverse power-law frequency dependence with unspecified index) -- suggestive of highly-correlated, possibly self-organised critical, behaviour -- have been observed in diverse accretion systems (reviewed by, for example, Dendy et al. \cite{dendy}). It is also well established that instabilities occuring in accretion disks (Dubus et al. \cite{dubus} and references therein) can give rise to anomalous statistics. However, these statistics will differ from those of fully-developed turbulence, and hence we seek to constrain such models using the differencing and rescaling technique described below.

Self-similarity, non-Gaussianity and non-trivial temporal scaling together are strong indications of highly-correlated processes such as turbulence (Bohr et al. \cite{bohr}). We will show how trivial scaling of near-Gaussian fluctuations in the Crab X-ray signal -- evidence of diffusive transport -- contrasts with non-trivial scaling of non-Gaussian fluctuations in the X-ray signals from Cygnus X-1 and GRS1915+105. Whilst there may be other methods we could apply, we are confident that the methods described here are sufficiently insensitive to the timing and counting errors in the data; many techniques are unreliable even when random errors are small (Sornette \cite{sorn}).

\section{Techniques}

\subsection{The Probability Density Function (PDF)}\label{PDF}
The first step in our analysis of each data set is the construction of its PDF. The PDF $P(y)$ of a variable $Y$ is defined such that the probability that $Y$ lies within a small interval $dy$ centred on $Y=y$, is equal to $P(y)dy$. $P(y)$ is normalised so that 
\[\int_{y_{min}}^{y_{max}}P(y)dy=1.\]
Here, we use $dy=(y_{max}-y_{min})/100$ and further normalise each PDF to its mean $\langle y\rangle$ and standard deviation $\sigma$ to enable comparison with theoretical distributions. We note that the PDF was used by Antar et al. (\cite{antar}) to characterise turbulent fluctuations in tokamak edge plasmas, and by Bhavsar \& Barrow (\cite{bhav}) to model the magnitude distribution of the brightest cluster galaxies. Bramwell et al. (\cite{bram1,bram2,bram3}) discuss the PDFs of fluctuations in highly-correlated systems (see Sect.~\ref{stats}), and Burlaga (\cite{burlaga}) presents a review of log-normal distributions in the turbulent solar wind.   

\subsection{Growth of range}\label{correlation}
Consider a self-similar function $y\left(t\right)$. The difference between the maximum and minimum values of $y$ during a time interval $\Delta t$ defines its range for that interval, $R\left(\Delta t\right)$. Since $y$ is self-similar, the ensemble-averaged value of $R$ will scale with $\Delta t$. We may write
\begin{equation}
\left<R\left(\Delta t\right)\right>=c\Delta t^{H}\label{range}
\end{equation}
with $c$ and $H$ constants; $H$ here defines the Hurst exponent (Hastings \& Sugihara \cite{frac}). For data that are only approximately self-similar, we use the above relation to check their closeness to self-similarity and also to obtain an effective value for $H$ as follows. By moving the window $\Delta t$ one point at a time through the raw data, an array of $R\left(\Delta t\right)$ values is created from which the mean $\left<R\right>$ is found (reducing the effects of uneven sampling). This is repeated for a range of $\Delta t$ within the length of the data set. A plot of $\log \left<R\left(\Delta t\right)\right>$ against $\log \Delta t$ will reveal any deviations from self-similarity whilst the slope will give the best estimate of $H$. We use linear regression to calculate the $95\%$ confidence interval for $H$ (see Figs.~\ref{gcr}, \ref{gcy} and \ref{gg}).  

Trivially, a function that is exactly constant over time has $H=0$. At the other extreme, $H=1$ indicates a function whose range increases linearly with time (for positive $c$ in Eq.~\ref{range}). Intermediate values of $H$ are generated by fractal functions, random Gaussian noise ($H\approx 0.2$), and Gaussian random walks (whose next value in time is the sum of the previous value and a random Gaussian increment; $H\approx 0.5$). The value of $H$ does not uniquely establish correlation, however; uncorrelated series may present significant probabilities of observing greater values as the timescale increases. Consequently, the growth of range can be rather insensitive as a measure of correlation. We can in principle define a measure of correlation $\beta$ in terms of fractal exponents such as $H$ ($\beta =2H+1$). Following Malamud \& Turcotte (1999), $\beta=0$ for uncorrelated noise and $\beta=2$ for a Gaussian random walk. However, the use of only one method to estimate an unknown exponent (and hence $\beta$) is to be avoided (Schmittbuhl et al. \cite{schmitt}). For example, comparing values of $H$ obtained from the growth of range of synthesized series with known $\beta$,  Malamud \& Turcotte (\cite{mala}) find that $H\approx 0.2$ for $\beta< 0.5$, whereas the slopes of the power spectra are equal to $-\beta$ for $0<\beta<2$. We therefore obtain, for comparison, estimates of the strengths of correlation from the slopes of power spectra with Thompson multi-tapering (Percival \& Walden \cite{percival}) to reduce the variance.
  
The Hurst exponent has been used to quantify solar magnetic complexity (Adams et al. \cite{adams}), correlation in the dynamics of the upper photosphere (Hanslmeier et al. \cite{hans}), and persistence in solar activity (Lepreti et al. \cite{lepreti} and references therein). In conjunction with rescaling techniques, the Hurst exponent has also been used to quantify self-similarity and long-range correlations in turbulent fluctuations in magnetic fusion plasmas (Carreras et al. \cite{carr}). 

\subsection{Differencing and rescaling}
\begin{figure*}
   
	\includegraphics[width=17cm]{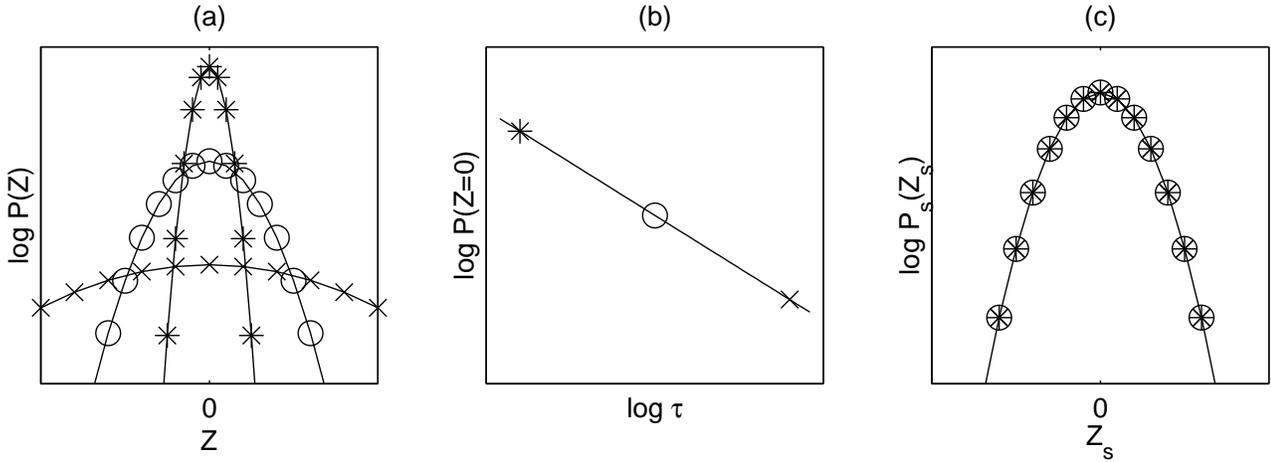}
\caption{(a) Unscaled PDFs of differenced series (Eq.~\ref{differ}), drawn from a Gaussian random walk $y(t)$, each curve for a different $\tau$. (b) Power-law scaling of $P(0,\tau)$ with $\tau$; -1/slope yields exponent $\alpha$. (c) Rescaled PDFs (Eq.~\ref{rescale}) share the same (Gaussian) curve for all $\tau$.}\label{steps}
\end{figure*}

Starting from the raw data $y\left(t\right)$, we first form a set of differenced series $Z\left(t,\tau\right)$ for a range of values of the time-lag $\tau$:
\begin{equation}
Z\left(t,\tau\right)=y\left(t\right)-y\left(t-\tau\right).\label{differ}
\end{equation}
From these we calculate a set of PDFs for $Z(t)$, one for each value of $\tau$, which we denote by $P\left(Z,\tau\right)$ (not normalised) -- see Fig.~\ref{steps}a. It is often found to be the case that the peaks $P\left(0,\tau\right)$ scale approximately as $\tau ^{-\alpha}$, as in Fig.~\ref{steps}b. Such scaling is, of course, sought using the values of $P\left(0,\tau\right)$ because they are the most accurate points of the PDFs, having the most data points in their bins. The question then arises whether the differenced series $Z\left(t,\tau\right)$ can all be derived from a single PDF that scales. That is, whether a common functional form emerges if we use the exponent $\alpha$ to rescale the axes such that 
\begin{equation}
Z\rightarrow Z\tau ^{\alpha}=Z_{s}\hspace{5mm}\mathrm{and}\hspace{5mm}P\rightarrow P\tau ^{-\alpha}=P_{s}\label{rescale},
\end{equation}
causing the separate PDFs to collapse onto the same ($\tau=1$) curve  -- Fig.~\ref{steps}c (compare Mantegna \& Stanley \cite{diff}).
Clearly, a reasonable straight line in Fig.~\ref{steps}b (at least up to some maximum timescale $\tau_{max}$) is necessary but not sufficient for rescaling to succeed, since only the peaks of $P\left(Z,\tau\right)$ are used to determine $\tau$. Physically, successful rescaling of the differenced X-ray data would imply, on all timescales up to $\tau_{max}$, that (i) the sizes of X-ray fluctuations (that is, the \emph{differences} between the observed values) are governed by a single type of process, and (ii) the total X-ray output is correlated, not random, in time. In the trivial case where $P\left(Z,\tau\right)$ is independent of $\tau$, we would infer the absence of temporal correlation in the X-ray output.

The value of $\alpha$ is given by $\alpha=-1/m$ where $m$ is the slope of $log P(0)$ v. $log \tau$ as shown in Fig.~4(b). $\alpha$ characterises the common functional form of the distributions $P(Z)$ viz:

\begin{enumerate}
	\item $0<\alpha<2$ L\'{e}vy (power-law with $\sigma\rightarrow\infty$)
	\item $\alpha=2$   Gaussian (finite $\sigma$)
	\item $\alpha>2$   power-law with finite $\sigma$
\end{enumerate} 
where $P(Z)\sim |Z|^{-(1+\alpha)}$ for $Z\rightarrow\pm\infty$ (Sornette \cite{sorn}).

Thus the differencing and rescaling procedure not only reveals the timescales over which physical processes occur, but can also confirm any correlation suggested by the growth of range and inverse power-law form of the power spectra. Moreover, it quantifies the asymptotic behaviour of the distribution of fluctuations, which is essential for constraining turbulence/instability models (Bohr et al. \cite{bohr}).
 
Differencing was used by Mantegna \& Stanley (\cite{diff}) to investigate fluctuations in the value of a financial index, but neither this technique nor the growth of range has yet (to our knowledge) been applied to astrophysical X-ray sources.

\section{Results}
 
\subsection{The Crab}
\begin{figure}

	\resizebox{\hsize}{!}{\includegraphics{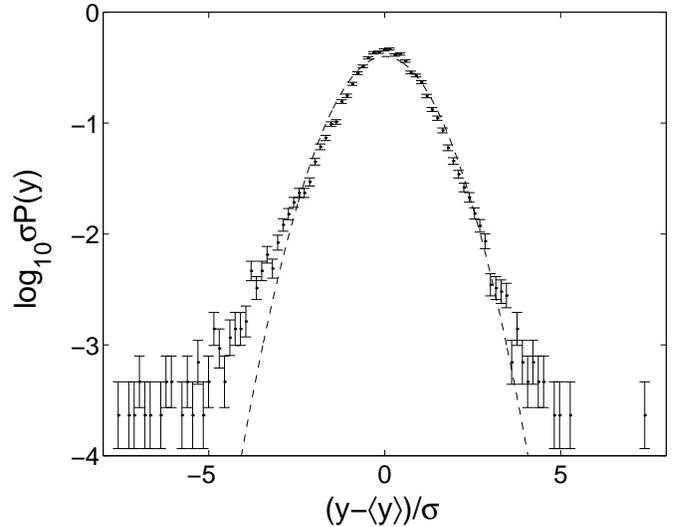}}
	\caption{PDF for Crab X-ray time-series, normalised with respect to $\langle y\rangle$ and $\sigma$; dashed line is Gaussian.}\label{pcr}
\end{figure}

\begin{figure}

	\resizebox{\hsize}{!}{\includegraphics{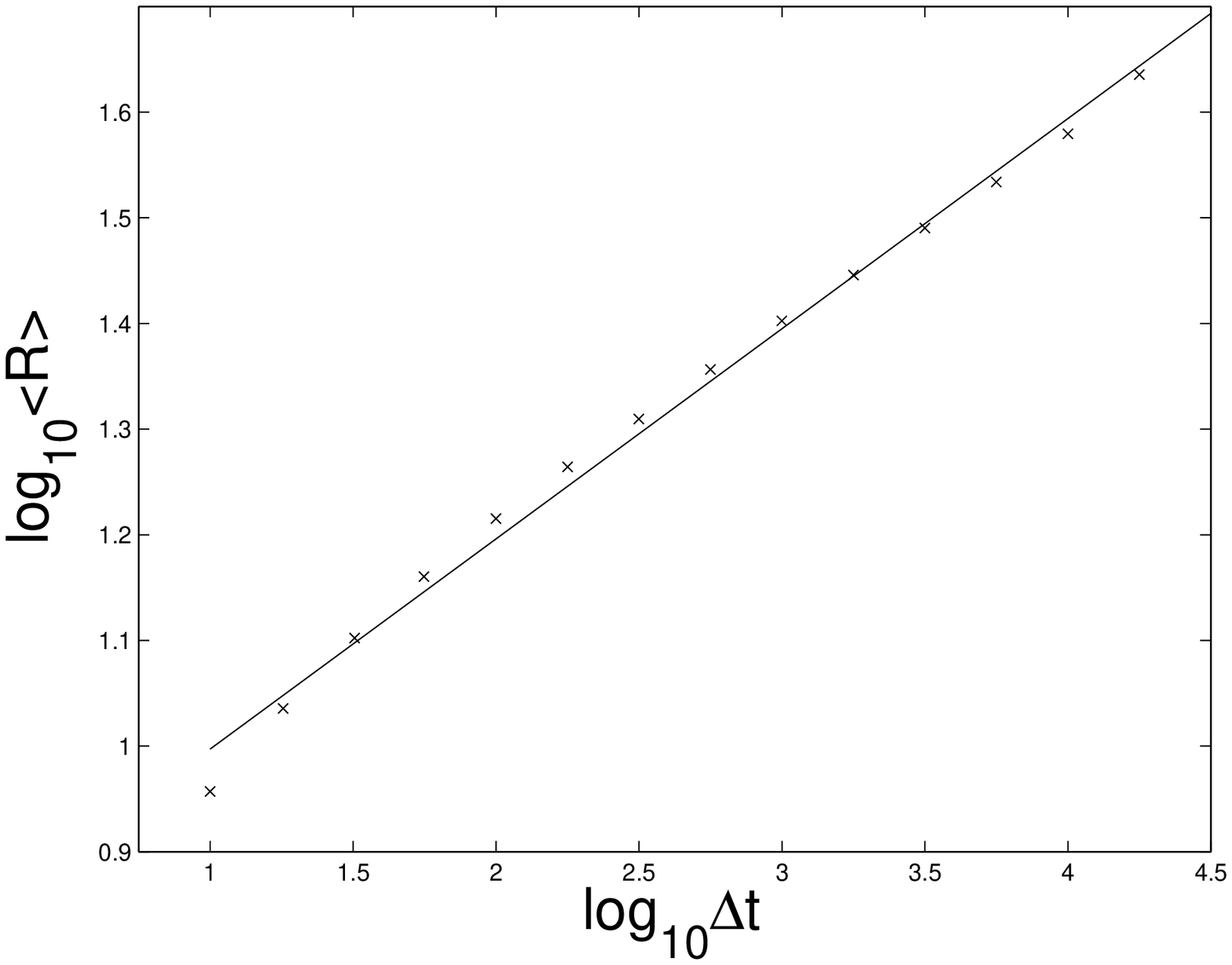}}
	\caption{Logarithmic plot of growth of range $R(\Delta t)$ (Eq.~\ref{range}) with $\Delta t$ for Crab X-ray time- series; slope gives Hurst exponent $0.19<H<0.21$.}\label{gcr}
\end{figure}

\begin{figure}

	\resizebox{\hsize}{!}{\includegraphics{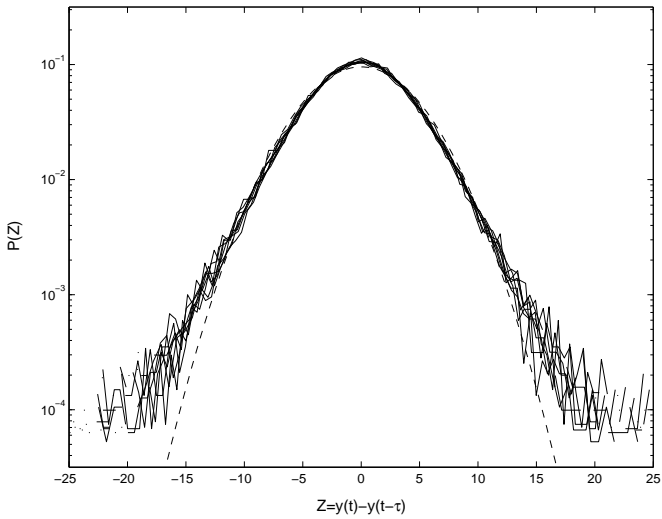}}
	\caption{Unscaled PDFs of differenced time-series (Eq.~\ref{differ}) for Crab, with parameter $\tau$ stepping up in half-integer powers of timesteps, to maximum $10^{4}$; dashed line is Gaussian.}\label{dcr}
\end{figure}

Fig.~\ref{pcr} shows the PDF of the total X-ray count-rate $y\left(t\right)$ that form the time-series for the Crab (Fig.~\ref{lccr}). 
The curve is close to Gaussian but with longer tails. Growth of the range is as low as that of Gaussian noise ($H\approx 0.2$, Fig.~\ref{gcr}); the slope of the power spectrum is better at detecting weak correlation (see Sect.~\ref{correlation}) and gives $0.26\leq\beta\leq 0.28$ at the $95\%$ confidence level.
Interestingly, the PDFs of the differenced data (Fig.~\ref{dcr}) require no rescaling.
That is, not only is the type of distribution independent of $\tau$, but the differences themselves show no spread over time. We infer that the raw time-series is uncorrelated; if the correlation suggested by the power spectrum exists, it is too weak to be detected by this method.

\subsection{Cygnus X-1}
\begin{figure}
	
	\resizebox{\hsize}{!}{\includegraphics{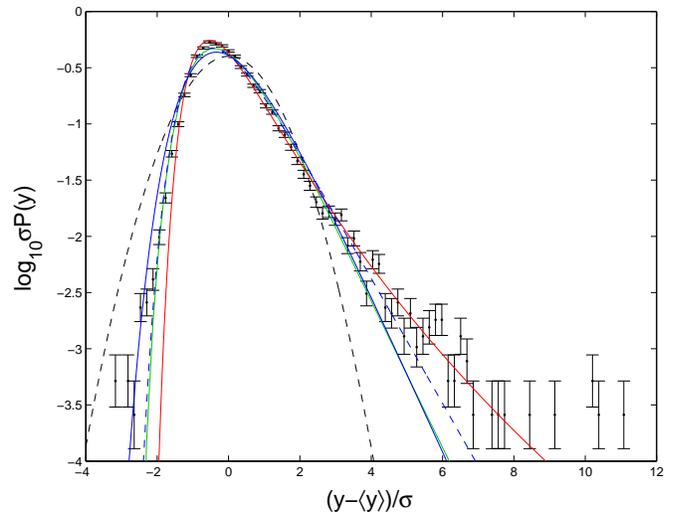}}
	\caption{PDF for Cygnus X-1 X-ray time-series, normalised with respect to $\langle y\rangle$ and $\sigma$; black dashed line Gaussian, green solid line log-normal, red solid line Fr\'{e}chet $a=1.1$, blue dashed line Gumbel $a=1$, blue solid line Gumbel $a=2$.}\label{pcy}
\end{figure}

Fig.~\ref{pcy} shows the PDF of the raw count-rates from Cygnus X-1 (Fig.~\ref{lccy}) over a continuous, relatively quiescent three-year period.
In contrast to Fig.~\ref{pcr}, this PDF is clearly non-Gaussian. It appears possible to characterise some aspects of its non-Gaussian statistical properties in terms of, for example, the Gumbel and Fr\'{e}chet distributions whose properties we outline in Sect.~\ref{stats}. Meanwhile, we note from Fig.~\ref{pcy} that the distribution of small amplitude events appears to fit the left-hand tail of Gumbel distributions with $a=[1,2]$ while the distribution of large amplitude events appears to fit the right-hand tail of a Fr\'{e}chet distribution with $a=1.1$. This suggests that the total flux has contributions from different physical processes, perhaps arising in different parts of the accretion disk and its surroundings, whose individual PDFs could differ. Fitting to each of the many component distributions is problematic; here we simply note that the PDF in Fig.~\ref{pcy} could not be reproduced by summing Gaussian PDFs.  

\begin{figure}

	\resizebox{\hsize}{!}{\includegraphics{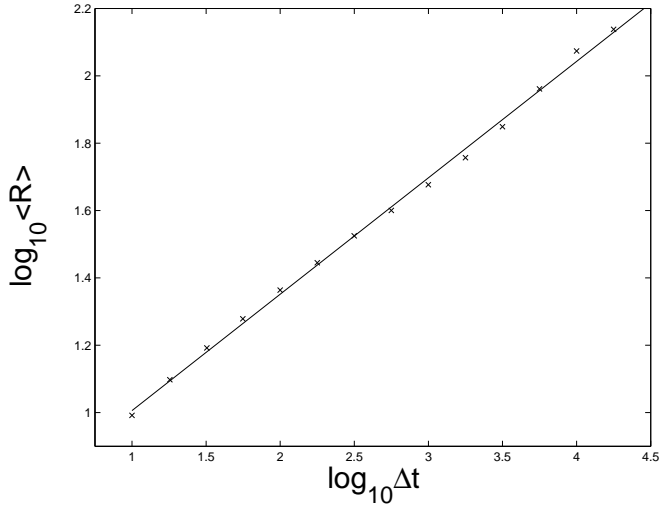}}
	\caption{Logarithmic plot of growth of range $R(\Delta t)$ (Eq.~\ref{range}) with $\Delta t$ for Cygnus X-1 X-ray time-series; slope gives Hurst exponent $0.34<H<0.36$.}\label{gcy}
\end{figure}

\begin{figure}

	\resizebox{\hsize}{!}{\includegraphics{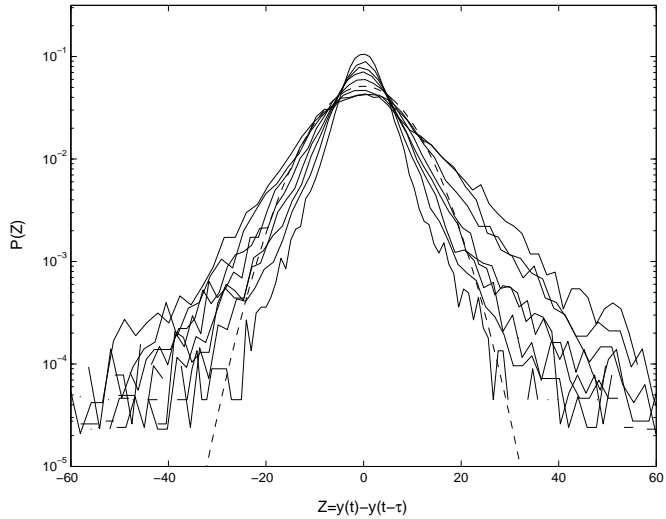}}
	\caption{Unscaled PDFs of differenced time-series (Eq.~\ref{differ}) for Cygnus X-1, with parameter $\tau$ stepping up in half-integer powers of timesteps, to maximum $10^{4}$. Curves with lower $P(0)$ and broader tails correspond to higher values of $\tau$.}\label{ucy}
\end{figure}

\begin{figure}

	\resizebox{\hsize}{!}{\includegraphics{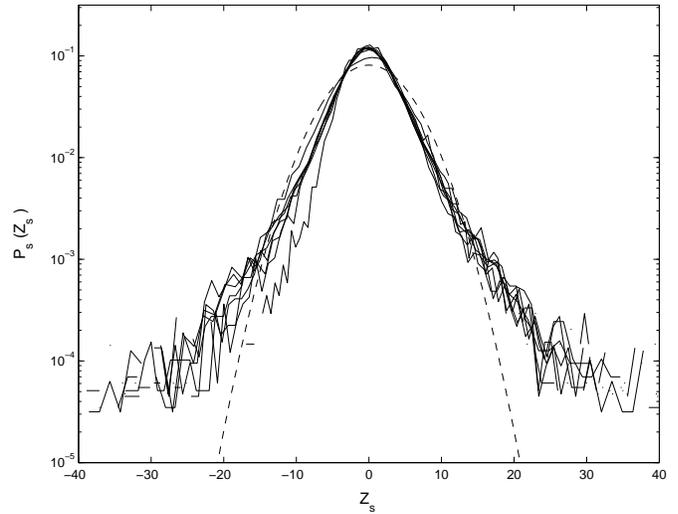}}
	\caption{Rescaled PDFs of differenced time-series for Cygnus X-1, with parameter $\tau$ stepping up in half-integer powers of timesteps, to maximum $10^{4}$; dashed line is Gaussian. Here, following Eq.~\ref{rescale}, $Z_{s}$ denotes the differences rescaled according to $\tau$ and $\alpha$, where $\alpha$ is obtained from values of $P(0,\tau)$ in Fig.~\ref{ucy} (compare Figs.~\ref{steps}a and \ref{steps}b); $5.4\leq\alpha\leq 9.9$.}\label{scy}
\end{figure}

Cygnus X-1 has a substantially higher growth of range ($H\approx 0.35$, Fig.~\ref{gcy}) than the Crab, but still below that of a Gaussian random walk ($H\approx 0.5$). This is confirmed by the power spectrum whose slope gives $1.00\leq\beta\leq 1.02$ at the $95\%$ confidence level (see Sect.~\ref{correlation}).
This higher growth is evident in the unscaled PDFs of the differenced Cygnus X-1 data in Fig.~\ref{ucy}.
By deriving a scaling exponent $\alpha$ from the peaks of Fig.~\ref{ucy}, and using it to rescale as in Fig.~\ref{scy}, it is clear that the increments scale remarkably well over the full range of the data (four decades in time). This establishes both the existence of correlation in the X-ray output, and variations controlled by one type of process, on timescales up to three years.
However, with its slightly higher peak and broader tails, the PDF is distinctly non-Gaussian. In fact, the raw X-ray time-series of Cygnus X-1 is correctly described as a weakly leptokurtic random walk; that is, the PDF of the increments is long-tailed (Bouchard \& Potters \cite{fin}). The asymptotic form of the increments  $P(Z)$ is quantified by $\alpha$; in this case, $5.4\leq\alpha\leq 9.9$ with $95\%$ confidence.

\subsection{GRS1915+105}
\begin{figure}

	\resizebox{\hsize}{!}{\includegraphics{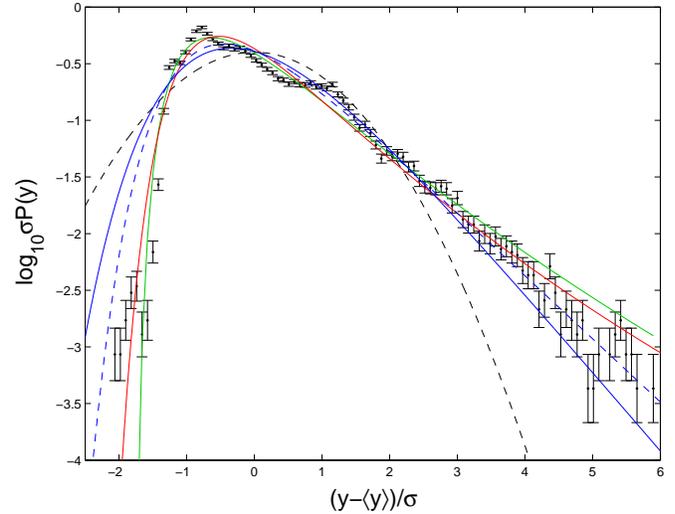}}
	\caption{PDF for GRS1915+105 X-ray time-series, normalised with respect to $\langle y\rangle$ and $\sigma$; black dashed line Gaussian, green solid line log-normal, red solid line Fr\'{e}chet $a=1.1$, blue dashed line Gumbel $a=1$, blue solid line Gumbel $a=2$.}\label{pg}
\end{figure}

\begin{figure}

	\resizebox{\hsize}{!}{\includegraphics{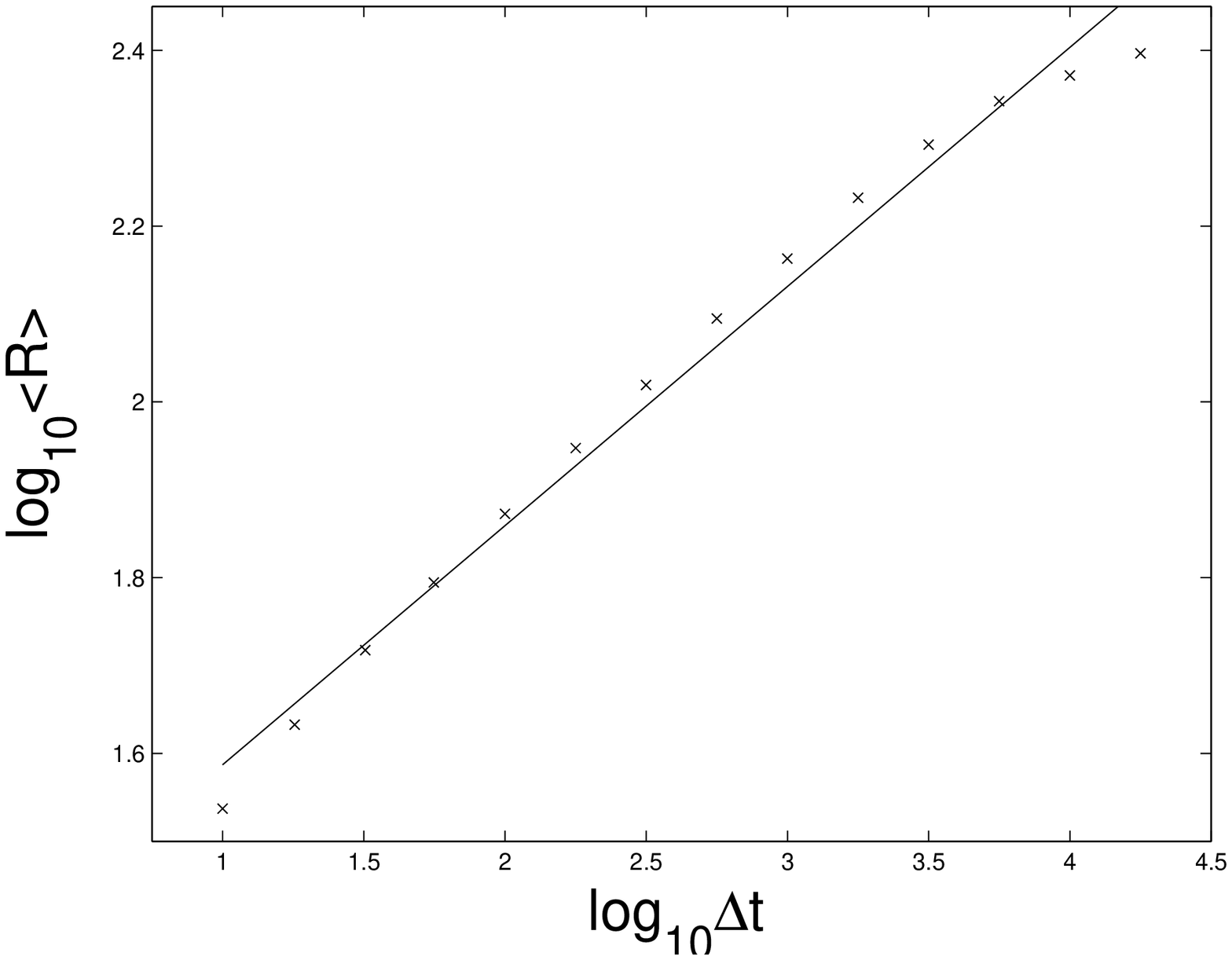}}
	\caption{Logarithmic plot of growth of range $R(\Delta t)$ (Eq.~\ref{range}) with $\Delta t$ for GRS1915+105 X-ray time-series; slope gives Hurst exponent $0.25<H<0.29$.}\label{gg}
\end{figure}

\begin{figure}

	\resizebox{\hsize}{!}{\includegraphics{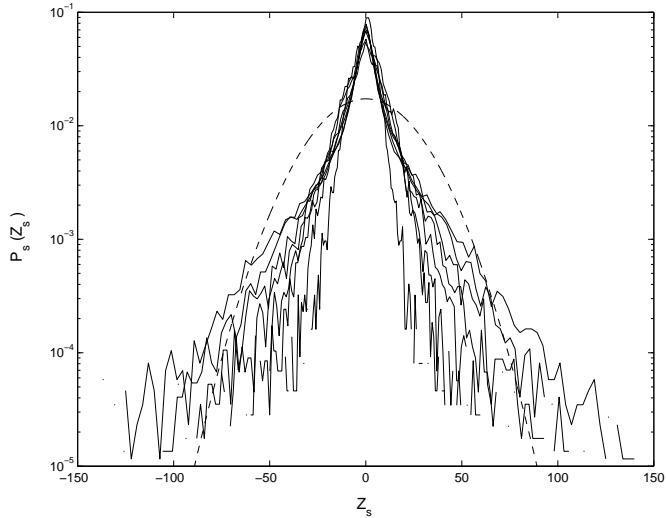}}
	\caption{Rescaled PDFs of differenced X-ray time-series for GRS1915+105, with parameter $\tau$ stepping up in half-integer powers of timesteps, to maximum $10^{4}$; dashed line is Gaussian. Here, following Eq.~\ref{rescale}, $Z_{s}$ denotes the differences rescaled according to $\tau$ and $\alpha$, where $\alpha$ is obtained from values of $P(0,\tau)$ (compare Figs.~\ref{steps}a and \ref{steps}b). Curves with lower $P_{s}(0)$ and broader tails correspond to higher values of $\tau$.}\label{sg}
\end{figure}

\begin{figure}

	\resizebox{\hsize}{!}{\includegraphics{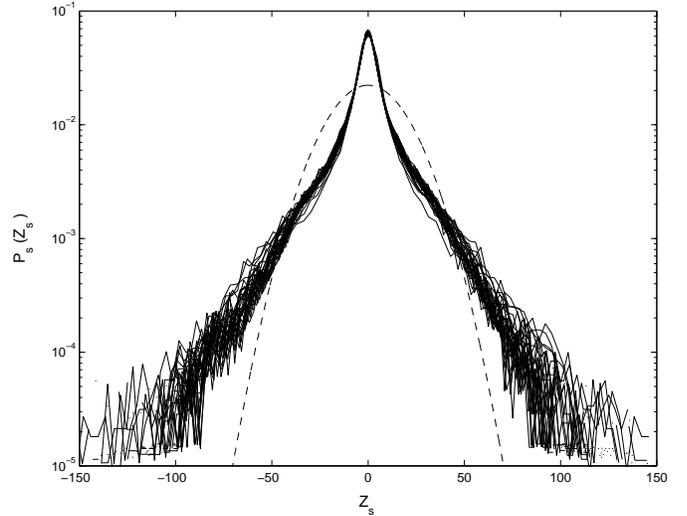}}
	\caption{Rescaled PDFs of differenced GRS1915+105 X-ray time-series, with $\tau$ stepping up in single integer numbers of timesteps, to a maximum value $10^{1.5}$; dashed line is Gaussian. $5.5\leq\alpha\leq 6.3$.}\label{s2g}
\end{figure}

Fig.~\ref{pg} shows the PDF of the raw data from GRS1915+105 (Fig.~\ref{lcgr}).
Like Cygnus X-1, GRS1915+105 is better described in terms of non-Gaussian statistics than Gaussian. For example, the extreme left- and right-hand tails fit Fr\'{e}chet and Gumbel curves respectively, the reverse situation of Cygnus X-1. Note also that there are several peaks, suggesting a multi-component source involving at least two physical processes with different statistical properties. With $H\approx 0.3$ (Fig.~\ref{gg}), midway between Gaussian noise ($H\approx 0.2$) and a Gaussian random walk ($H\approx 0.5$), the growth of range of GRS1915's count-rate lies between those of the Crab and Cygnus X-1. This is confirmed by the slope of the power spectrum: $0.91\leq\beta\leq 0.93$ at the $95\%$ confidence level (see Sect.~\ref{correlation}). 
The PDFs of the increments do not rescale over the full four decades, see Fig.~\ref{sg}.
The separate PDFs rescale approximately onto the same curve for $\tau$ up to only 1.5 decades (see Fig.~\ref{s2g}) with stronger leptokurtosis than for Cygnus X-1.
However, we cannot unambiguously identify scaling in this r\'{e}gime given the variation in sampling interval; see Appendix.
 
\section{Non-Gaussian statistical properties}\label{stats}
The tails of the PDFs of Cygnus X-1 and GRS1915+105 are clearly non-Gaussian. Here, we compare these PDFs with those arising from extremal statistics. The two limiting distributions of interest are ``Gumbel's asymptote'' and Fr\'{e}chet (Fisher \& Tippett \cite{fisher}; Gumbel \cite{gum}; Sornette \cite{sorn}). In outline, the limiting distributions result from selecting the maximum value $y_{max}$ from each of a large number of large samples whose individual members are drawn from a distribution $P(y)$. When $P(y)$ decreases more rapidly than any power-law (as $y\rightarrow\infty$), ``Gumbel's asymptote'' has the form
\begin{equation}
P_{G}(y_{max})=K(e^{u-e^{u}})^{a}\hspace{5mm}\mathrm{with}\hspace{5mm}u=b(y-s)\label{gumeq}
\end{equation}
where in the limit of an infinite number of measurements $a\equiv 1$; the constants $K$, $b$, and $s$ are fixed by normalisation as in Sect.~\ref{PDF}.

Fr\'{e}chet distributions $P_{F}(y_{max})$ arise in the same manner when the underlying PDF $P(y)$ is power-law. Mathematically, $P_{F}(y_{max})$ is defined by Eq.~\ref{gumeq} but with $u=\alpha+\beta\ln(1+y/G)$, where $K$, $b$, $s$, $\alpha$, and $G$ are again fixed by normalisation as in Sect.~\ref{PDF}, and $\beta=(1-a)^{-1}$. These curves exist for $1<a<3/2$.

Physically, the fact that the tails of the Cygnus X-1 and GRS1915+105 data can be fitted to Gumbel and Fr\'{e}chet distributions may suggest that the observed signals have the character of maximal events. In this case, they would be the brightest among multiple events (whose PDF could be power-law) occuring within each observational time window. Interestingly, extremal statistics in a global measure are found in turbulent fluids and other highly-correlated systems (Bramwell et al. \cite{bram1,bram2,bram3}). In these cases, anomalous values of $a$ are found for Gumbel's asymptote, and we plot $a=1$ and $2$ along with the Fr\'{e}chet curve $a=1.1$ for comparison with the X-ray data. Also, since our global quantity is emitted flux rather than absorbed power, our curves show the opposite handedness to the results of Bramwell et al. (\cite{bram1}).

We also plot the log-normal PDFs having the same $\langle y\rangle$ and $\sigma$ as the data, and these curves fit as closely as extremal distributions (see Figs.~\ref{pcy} and \ref{pg}). The significance and origin of extremal and log-normal distributions is currently of considerable interest in statistical physics and turbulence studies (Bramwell et al. \cite{bram3}; Burlaga \cite{burlaga}).    

\section{Conclusions}
Using three key statistical methods -- the PDF, the growth of range, and differencing and rescaling -- we have identified and quantified some fundamental contrasts in the character of the total X-ray output of the Crab, Cygnus X-1, and GRS1915+105 over multi-year time intervals. The Crab shows near-Gaussian behaviour with its low Hurst exponent and lack of temporal scaling. This is to be expected since the flux emanates from a very wide region over which correlation is unlikely, there being no observational evidence of an accretion disk in this source. In contrast, Cygnus X-1 has a time-series with scaling over three years, and resembles a random walk with leptokurtic increments whose PDF satisfies $P(Z)\sim |Z|^{-(1+\alpha)}$ for $Z\rightarrow\pm\infty$ and $\alpha\approx 8$. Its PDF closely follows multi-component Gumbel and Fr\'{e}chet curves, suggestive of the dominance of maximal events. Similarly, the output from GRS1915+105 lies close to such curves and its increments are also leptokurtic, although the roles of Fr\'{e}chet and Gumbel components in the tails are reversed. However, as suggested by a lower Hurst exponent, results from this source are consistent with only short-range scaling with $\alpha\approx 6$, and more data are required to establish this finding. Thus we have evidence that the two accreting objects display a degree of correlation in their X-ray time-series, which is absent from the nonaccreting Crab. This is a quantitative, observational, and model-independent measure of anomalous (non-diffusive) transport in accretion disks. 

\begin{acknowledgements}
We are grateful to John Kirk, Michel Tagger, and Nick Watkins for helpful suggestions. JG acknowledges a Research Studentship and SCC a Lecturer Fellowship from the UK Particle Physics and Astronomy Research Council. This work was also supported in part by the UK DTI and Euratom. S.C. acknowledges support from grant F/00-180/A from the Leverhulme Trust. Data provided by the ASM/RXTE teams at MIT and at the RXTE SOF and GOF at NASA's GSFC.
\end{acknowledgements}

\section*{Appendix: Confidence limits for $\tau$}
\restylefloat{table}
\begin{table}[H]
\begin{tabular}{l|ll|ll|ll} \hline 
$\log10(\tau)$ & \multicolumn{6}{l}{90\% confidence limits for $\log10(\tau)$} \\
    & \multicolumn{2}{l}{Crab} & \multicolumn{2}{l}{Cygnus X-1} & \multicolumn{2}{l}{GRS1915+105} \\
    & Min. & Max. & Min. & Max. & Min. & Max.\\ \hline
0.5 & -1.3 & 1.0 & -1.4 & 1.0 & -1.5 & 1.0 \\
1.0 & -0.2 & 1.5 & 0.4 & 1.4 & 0.3 & 1.5 \\
1.5 & 0.3 & 2.0 & 1.1 & 2.0 & 1.0 & 1.9 \\
2.0 & 1.0 & 2.4 & 1.6 & 2.4 & 1.6 & 2.4 \\
2.5 & 1.5 & 3.1 & 2.2 & 2.8 & 2.2 & 2.8 \\
3.0 & 2.3 & 3.4 & 2.8 & 3.2 & 2.8 & 3.2 \\
3.5 & 3.1 & 3.7 & 3.4 & 3.6 & 3.3 & 3.6 \\
4.0 & 3.9 & 4.0 & 3.9 & 4.0 & 3.9 & 4.0 \\ \hline 
\end{tabular}

\caption{Median-centred 90\% confidence limits for $\log10(\tau)$ in the X-ray time series of the Crab, Cygnus X-1, and GRS1915+105. When considering fluctuations over small numbers of data points, a consequence of the uneven observational sampling times is that a broad range of actual times are represented by each value of $\tau$  (see Fig.~\ref{s2g}). To quantify these ranges, this table lists the limits between which $90\%$ of the actual times lie for each value of $\tau$.}
\end{table}


\begin{thebibliography}{}
\bibitem[1997]{adams} Adams, M., Hathaway, D.~H., Stark, B.~A., \& Musielak, Z.~E. 1997, Sol. Phys., 174, 1, 341
\bibitem[2001]{antar} Antar, G.~Y., Devynck, P., Garbet, X., \& Luckhardt, S.~C. 2001, Phys. Plasmas, 8, 5, 1612
\bibitem[1991]{balb} Balbus, S.~A., \& Hawley, J.~F. 1991, ApJ, 376, 214
\bibitem[2000]{belloni} Belloni, T., Klein-Wolt, M., M\'{e}ndez, M., van der Klis, M., \& van Paradijs, J. 2000, A\&A, 355, 271
\bibitem[2001]{bhav} Bhavsar, S.~P., \& Barrow, J.~D. 1985, MNRAS, 213, 857
\bibitem[1998]{bohr} Bohr, T., Jensen, M.~H., Paladin, G., \& Vulpiani, A. 1998, Dynamical Systems Approach to Turbulence (Cambridge University Press, Cambridge)
\bibitem[2000]{fin} Bouchard, J., \& Potters, M. 2000, Theory of Financial Risk: from Statistical Physics to Risk Management (Cambridge University Press, Cambridge)
\bibitem[2001]{asm} Bradt, H.~V., Chakrabarty, D., Cui, W. et al. 2001, ASM Light Curves Overview (ASM/RXTE team, Massachusettes Institute of Technology)
\bibitem[1998]{bram1} Bramwell, S.~T., Holdsworth, P.~C.~W., Pinton, J.-F. 1998, Nature, 396, 552
\bibitem[2000]{bram2} Bramwell, S.~T., Christensen, K., Fortin, J.-Y. et al. 2000, PRL, 84, 17, 3744
\bibitem[2001]{bram3} Bramwell, S.~T., Fortin, J.-Y., Holdsworth, P.~C.~W. et al. 2001, PRE, 63, 4
\bibitem[2001]{burlaga} Burlaga, L.~F. 2001, JGR, 106, A8, 15917
\bibitem[1998]{carr} Carreras, B.~A., van Milliger, B., Pedrosa, M.~A. et al. 1998, PRL, 80, 4438     
\bibitem[2001]{chaty} Chaty, S., Rodr\'{\i}guez, L.~F., Mirabel, I.~F. et al. 2001, A\&A, 366, 1035 
\bibitem[1998]{dendy} Dendy, R.~O., Helander, P., \& Tagger, M. 1998, A\&A, 337, 962
\bibitem[2000]{dhawan} Dhawan, V., Mirabel, I.~F., \& Rodr\'{\i}guez, L.~F. 2000, ApJ, 543, 373
\bibitem[2001]{dubus} Dubus, G., Hameury, J.-M., \& Lasota, J.-P. 2001, A\&A, 373, 251
\bibitem[1928]{fisher} Fisher, R.~A., \& Tippett, L.~H.~C. 1928, Proc. Camb. Phil. Soc., XXIV, 180
\bibitem[1998]{gamm} Gammie, C.~F. 1998, Proc. Eighth Ann. Astrophys. Conf. in Maryland (AIP, New York)
\bibitem[1958]{gum} Gumbel, E.~J. 1958, Statistics of Extremes (Columbia University Press, New York)
\bibitem[2000]{hans} Hanslmeier, A., Kucera, A., Ryb\'{a}k, J., Neunteufel, B. \& W\"{o}hl, H. 2000, A\&A, 356, 308
\bibitem[1993]{frac} Hastings, H.~M., \& Sugihara, G. 1993, Fractals: a User's Guide for the Natural Sciences (Oxford University Press, New York)
\bibitem[2000]{lepreti} Lepreti, F., Fanello, P.~C., Zaccaro, F., \& Carbone, V. 2000, Sol. Phys., 197, 1, 149
\bibitem[2000]{cy} Maccarone, T.~J., Coppi, P.~S., \& Poutanen, J. 2000, ApJ, 537, L107
\bibitem[1999]{mala} Malamud, B.~D., \& Turcotte, D.~L. 1999, Advances in Geophysics, Vol.~40 (Academic Press, San Diego) 
\bibitem[1995]{diff} Mantegna, R.~N., \& Stanley, H.~E. 1995, Nature, 376, 46
\bibitem[1999]{mirabel} Mirabel, I.~F., \& Rodr\'{\i}guez, L.~F. 1999, Annu. Rev. Astron. Astrophys., 37, 409 
\bibitem[2000]{nay} Nayakshin, S., Rappaport, S., \& Melia, F. 2000, ApJ, 535, 798
\bibitem[1993]{percival} Percival, D.~B., \& Walden, A.~T. 1993, Spectral Analysis for Physical Applications: Multitaper and Conventional Univariate Techniques (Cambridge Universtity Press, Cambridge)
\bibitem[1972]{pringle} Pringle, J.~E., \& Rees, M.~J. 1972, A\&A, 21, 1
\bibitem[2000]{rao} Rao, A.~R., Yadav, J.~S., \& Paul, B. 2000, ApJ, 544, 1, 443
\bibitem[1995]{schmitt} Schmittbuhl, J., Vilotte, J.-P., \& Roux, S. 1995, PRE, 51, 131
\bibitem[1973]{shak} Shakura, N.~I., \& Sunyaev, R.~A. 1973, A\&A, 24, 337
\bibitem[2000]{sorn} Sornette, D. 2000, Critical Phenomena in Natural Sciences (Springer-Verlag, Berlin)
\bibitem[2001]{rxte} Swank, J.~H., Smale, H.~P., Boyd, P.~T. et al. 2001, RXTE Guest Observer Facility (GOF) (RXTE GOF, GSFC)
\bibitem[2000]{cr} Weisskopf, M.~C., Hester, J.~J., Tennant, A.~F. et al. 2000, ApJ, 536, L81   
\end{thebibliography}
\end{document}